# Alternative interdiffusion theory of manycomponent alloys

A.V. Nazarov[a,b,*], D. A. Belobraga[a], A.P. Melnikov[a]

[a] National Research Nuclear University MEPhI, (Moscow Engineering Physics Institute), 31, Kashirskoe shosse 115409, Moscow, Russia.

[b] Institute for Theoretical and Experimental Physics named by A.I. Alikhanov of NRC "Kurchatov Institute", 25 Bolshaya Cheremushkinskaya str.,117218, Moscow, Russia

**Abstract**

We examine the interdiffusion in multicomponent systems using the approach analogically to one developed earlier for description of interdiffusion in binary alloys. This approach in opposite to traditional theory, takes into consideration an active role of vacancies, equilibrium distribution of which is not supposed. In this case, in equations for flux components there are contributions conditioned by vacancy concentration gradient. As the vacancy diffusion coefficient is much larger than those for the components, a counteraction to this process will equalize the fluxes of components and, consequently, hinder from increase of deviation of vacancy concentration from the equilibrium one. If we substitute the expressions for fluxes in the equation of a continuity then we have the system of diffusion equations for components and vacancies. This system is solved and we have found a relation between the interdiffusion coefficients and the corresponding tracer diffusion coefficients. Interdiffusion coefficient equations sufficiently differ from traditional one (Darken's approach).

The analysis of the possible reasons of "sluggish" diffusion in multicomponent alloys is carried out on the basis of the derived equations for fluxes.

## 1. Introduction

Diffusion processes in metals and alloys play an important role in manufacturing technology and heat treatment and determining a lot of properties of these materials [1-6]. And besides interdiffusion that are basic in binary and multicomponent systems in the presence of spatial heterogeneity of chemical composition, are of particular interest to many technology processes. Interdiffusion underlies such important processes as homogenization of alloys, the application of protective coatings on the surfaces of materials, welding, the formation of the structure of multiphase materials, the growth and dissolution of precipitates and some other. Much attention is paid to high-entropy multicomponent alloys [7-9] due to some interesting properties and, in particular, their resistance to irradiation by high-energy particles. The special interest of many researchers is associated with slow diffusion in such alloys and an understanding of the causes of this effect.

In this paper approach that developed for binary system [10-16] is generalized for the multicomponent alloys. Remind that in this approach unlike in the Darken theory non-equilibrium vacancies are explicitly taken into account and their role in flattening of the fluxes of the components is clarified. What is more, a new version of the linearization of the equations of the original system is developed, allowing to find the terms slowly varying with time in the equations for the component concentrations. As shown before [11-13], exactly these terms give main contribution in concentration profile in the diffusion zone. Note that the equations for the fluxes of components in multicomponent systems are obtained for the first time in the within of the alternative theory of interdiffusion (ATID).

In the first section of the paper the mathematical apparatus of the approach is illustrated by the example of a binary system. In the second, the main stages of obtaining equations for fluxes in multicomponent systems are presented and the resulting equations are written out. The third section presents the results of illustrative calculations for the concentration dependences of the matrix of interdiffusioncoefficients and numerical solutions for a system of nonlinear diffusion equations for various values of the parameters. In the last part possible reasons of sluggish diffusion in high entropy alloys are discussed on the basis of the derived equations for fluxes.

## 2. Analysis of interdiffusion in binary alloys

According to works [4] equations for fluxes in binary alloys have the form when gradients of components are directed along axis X:

$$\vec{j} = -\frac{\gamma a^2}{\Omega}\left[(c_A\Gamma_A + c_B\Gamma_B)\frac{\partial c_v}{\partial x} - c_v\Gamma_A\frac{\partial c_A}{\partial x} - c_v\Gamma_B\frac{\partial c_B}{\partial x} - c_v c_A\frac{\partial \Gamma_A}{\partial x} - c_v c_B\frac{\partial \Gamma_B}{\partial x}\right], \quad (1)$$

$$\vec{J}_A = -\frac{\gamma a^2}{\Omega}\left[c_v\Gamma_A\frac{\partial c_A}{\partial x} - c_A\Gamma_A\frac{\partial c_v}{\partial x} - c_v c_A\frac{\partial \Gamma_A}{\partial x}\right], \quad (2)$$

$$\vec{J}_B = -\frac{\gamma a^2}{\Omega}\left[c_v\Gamma_B\frac{\partial c_B}{\partial x} - c_B\Gamma_B\frac{\partial c_v}{\partial x} - c_v c_B\frac{\partial \Gamma_B}{\partial x}\right], \quad (3)$$

where $c_A, c_B$ are the atom concentrations of kind A and B, $c_v$ is the vacancy concentration, $\Omega$ is the volume per lattice site,

$$\Gamma_A = \nu_A \exp\left(-\frac{Q_A}{kT}\right), \quad (4)$$

$\Gamma_A$ is the probability of the exchanging atoms of kind A with a vacancy per unit of time, $\nu_A$ is the attempt frequency, $Q_A$ is the potential barrier (activation energy); expression for $\Gamma_B$ is similar, $a$ is the lattice parameter, $\gamma$ is the dimensionless factor depending on the type of crystallographic structure and the geometry of diffusion atomic jumps into a vacancy.

Further, it is assumed that $\Gamma_A$ and $\Gamma_B$ are independent of $x, t$ to simplify mathematical transformations. The concentrations of components and vacancies for a we define as a function of $x, \xi$ and $t$ in contrast to [11]:

$$c_A(x + \xi, t), \quad c_B(x + \xi, t), \quad c_v(x + \xi, t) \text{ on range } 0 < \xi < l. \quad (5)$$

Underscore that within the interval there are no vacancy sinks and sources, but they may be on its boundaries. Further we represent each of the concentrations $c_A$, $c_B$, $c_v$ as the sum of two functions:

$$c_A(x + \xi, t) = c_A^0(x, t) + c_A^1(\xi, x, t), c_B(x + \xi, t) = c_B^0(x, t) + c_B^1(\xi, x, t),$$

$$c_v(x + \xi, t) = c^0(x, t) + c^1(\xi, x, t). \quad (6)$$

An example of such a presentation is the Taylor series. Since the Eqs. (1), (2), (3) for the fluxes are obtained under the assumption that the gradients are small, it is always possible to choose an interval $l$ such that the conditions are met:

$$\frac{c_A^1}{c_A^0} \ll 1, \quad \frac{c_B^1}{c_B^0} \ll 1, \quad \frac{c^1}{c^0} \ll 1. \quad (7)$$

Under these conditions, the continuity equations are valid not only for atoms, but also for vacancies:

$$\frac{\partial c_v}{\partial t} = -div\, \mathbf{J}_v,$$

$$\frac{\partial c_A}{\partial t} = -div\, \mathbf{J}_A, \quad (8)$$

$$\frac{\partial c_B}{\partial t} = -div\, \mathbf{J}_B.$$

Substituting the Eqs. (1), (2) for the fluxes into the continuity equations, we obtain the system of diffusion equations (since the equations for concentrations $c_B$ are similar to the equations for concentrations $c_A$, we don't write them down below).

The exact equations for $c_A$ и $c_v$ are:

$$\frac{\partial c_A}{\partial t} - A\frac{\partial}{\partial \xi}\left(c_v \frac{\partial c_A}{\partial \xi}\right) = -A\frac{\partial}{\partial \xi}\left(c_A \frac{\partial c_v}{\partial \xi}\right), \tag{9}$$

$$\frac{\partial c_v}{\partial t} - \frac{\partial}{\partial \xi}\left((Ac_A + Bc_B)\frac{\partial c_v}{\partial \xi}\right) = -A\frac{\partial}{\partial \xi}\left(c_v \frac{\partial c_A}{\partial \xi}\right) - B\frac{\partial}{\partial \xi}\left(c_v \frac{\partial c_B}{\partial \xi}\right), \tag{10}$$

we replace the linearized equations with subject to conditions (6):

$$\frac{\partial c_A}{\partial t} - Ac^0 \frac{\partial^2 c_A}{\partial \xi^2} = -Ac_A^0 \frac{\partial^2 c_v}{\partial \xi^2}, \tag{11}$$

$$\frac{\partial c_v}{\partial t} - D_V(t)\frac{\partial^2 c_v}{\partial \xi^2} = -D_A(t)\frac{\partial^2 c_A}{\partial \xi^2} - D_B(t)\frac{\partial^2 c_B}{\partial \xi^2}, \tag{12}$$

where

$$D_V(t) = Ac_A^0(x,t) + Bc_B^0(x,t), \quad D_A(t) \equiv Ac^0(x,t).$$

Note that in these equations x is a parameter.

We introduce the notation for the initial and boundary conditions:

$$\begin{aligned} u_A(\xi) &= c_A(\xi,0), \\ c_1^A(t) &= c_A(0,t), \quad c_2^A(\tau) = c_A(l,t), \\ v_0(\xi) &= c(\xi,0), \quad v_1(t) = c(0,t), \quad v_2(\tau) = c(l,t). \end{aligned} \tag{13}$$

The system of equations, apparently, has no analytical solution in the general case. However earlier [11-12] in the analysis of interdiffusion in binary alloys are shown that the solutions of the linearized system of equations consist of the sum of the terms that vary slowly and rapidly with time. Moreover, the component concentrations with accuracy of terms proportional to the concentration of vacancies depend only on slowly varying terms. Thus, if we search contributions to solutions that vary slowly with time, then the application of approach similar to the "reducing the description" is possible [17].

At first, we should find the solutions of the diffusion equations with the diffusion coefficient depending on time. The description of this approach is given in Appendix I.

Considering the form of solution obtained in Appendix I, the solutions of Eqs. (11) and (12) can be represented as series:

$$c_v(\xi,t) = \sum_{n=1}^{\infty} \varphi_n(t) \sin(\lambda_n \xi), \qquad c^A(\xi,t) = \sum_{n=1}^{\infty} \varphi_n^A(t) \sin(\lambda_n \xi), \tag{14}$$

were $\lambda_n = \pi n/l$.

Then

$$\frac{\partial^2 c_v}{\partial \xi^2} = -\sum_{n=1}^{\infty} \lambda_n^2 \varphi_n(t)\sin(\lambda_n \xi), \qquad \frac{\partial^2 c^A}{\partial \xi^2} = -\sum_{n=1}^{\infty} \lambda_n^2 \varphi_n^A(t)\sin(\lambda_n \xi). \tag{15}$$

And after simple transformations for $\varphi_n(t)$ and $\varphi_n^A(t)$ we get:

$$\varphi_n(t) = \varphi_n(0)\exp\left(-\lambda_n^2 \int_0^t D_V(t')dt'\right) + \frac{2\pi n}{l}\int_0^t \exp\left(-\lambda_n^2 \int_{t'}^t D_V(t'')dt''\right)v_{12}(t')D_V(t')dt' +$$

$$+ \int_0^t \exp\left(-\lambda_n^2 \int_{t'}^t D_V(t'')dt''\right)\lambda_n^2(D_A(t')\varphi_n^A(t') + D_B(t')\varphi_n^B(t'))dt', \tag{16}$$

$$v_{12}(t) = v_1(t) - (-1)^n v_2(t),$$

$$\varphi_n^A(t) = \varphi_n^A(0) \exp\left(-\lambda_n^2 \int_0^t D_A(t')t'\right) + \frac{2\pi n}{l^2} \int_0^t \exp\left(-\lambda_n^2 \int_{t'}^t D_A(t'')t''\right) c_{12}^A(t') D_A(t') dt' +$$

$$+ \int_0^t \exp\left(-\lambda_n^2 \int_{t'}^t D_A(t'')t''\right) \lambda_n^2 Ac^{0A}(t') \varphi_n(t') dt', \tag{17}$$

$$c_{12}^A(t) = c_1^A(t) - (-1)^n c_2^A(t),$$

and $\varphi_n^B$ has a similar form.

After substitution of $\varphi_n^A$ and $\varphi_n^B$ in Eq. (15) and transformations of the last term there, we obtain the solutions in the form of a series on the vacancy concentration (see Appendix I, Eqs. (I.12)). Restricting ourselves to the first order in this solution and retaining only slowly changing terms, we get:

$$\bar{\varphi}_n(t) = \bar{v}_{12}(t) + \frac{D_A(t)}{D_V(t)} \bar{\varphi}_n^A(t) + \frac{D_B(t)}{D_V(t)} \bar{\varphi}_n^B(t). \tag{18}$$

Substitution of Eq. (17) into the equation for the flux of kind A atoms gives:

$$j_A = -D_A \frac{\partial c_A}{\partial \xi} + Ac_A^0(t) \frac{\partial c}{\partial \xi} = -c_B^0 \frac{D_A D_B}{D_A c_A^0 + D_B c_B^0} \frac{\partial c_A}{\partial \xi} + c_A^0 \frac{D_A D_B}{D_A c_A^0 + D_B c_B^0} \frac{\partial c_B}{\partial \xi} +$$

$$+ Ac^{0A}(t) \frac{v_1(t) - v_2(t)}{l}. \tag{19}$$

Provided that the equilibrium concentration of vacancies for the current composition is maintained at the sinks, we have:

$$j_A = -\frac{D_A(t) D_B(t)}{D_A(t) c_A^0 + D_B(t) c_B^0} \frac{\partial c_A}{\partial \xi} + Ac_A^0 \frac{c_1^e - c_2^e}{l}, \tag{20}$$

where $c_e$ is the quasi-equilibrium vacancy concentration.

Since systems are close to ideal, the second term is small compared to the first one, but it should be taken into account in the equation for the flux of vacancies when estimating the shifting rate of marker displacement in the Kirkendall effect [16].

If we reduce the linearization interval $h$ and push it to zero, then:

$$\frac{\partial c_A}{\partial \xi} = \frac{\partial c_A}{\partial x}, \tag{21}$$

$$c_A(x + \xi, t) = c_A^0(x, t), \qquad c_B(x + \xi, t) = c_B^0(x, t),$$

and Eq. (20) takes a more familiar form:

$$j_A(x,t) = -\frac{D_A(x,t) D_B(x,t)}{D_A(x,t) c_A(x,t) + D_B(x,t) c_B(x,t)} \frac{\partial c_A}{\partial x}, \tag{22}$$

$$\widetilde{D} = \frac{D_A(x,t) D_B(x,t)}{D_A(x,t) c_A(x,t) + D_B(x,t) c_B(x,t)},$$

that is, the coefficient of interdiffusion has the same form as previously obtained [11]. However, now the coefficients $D_A(t)$ and $D_B(t)$ are time dependent and are not limited to any time interval. In other words, Eq. (22) is a renormalized equation for the flux of component A, moreover, the result of the renormalization is the consideration of the deviation of the vacancy concentration from the equilibrium in the equations for the fluxes. Note that the deviation leads to the flattening of the fluxes of the components in the binary alloy.

The dependences of the interdiffusion coefficients on the concentration for different ratios of the self-diffusion coefficients of A and B atoms are shown for comparison in Figure 1. These dependences are calculated by Eq. (22) and the Darken formula without taking into account the thermodynamic factor. The coefficients $D_A$ and $D_B$ do not depend on the concentrations and are constants for these illustrative calculations.

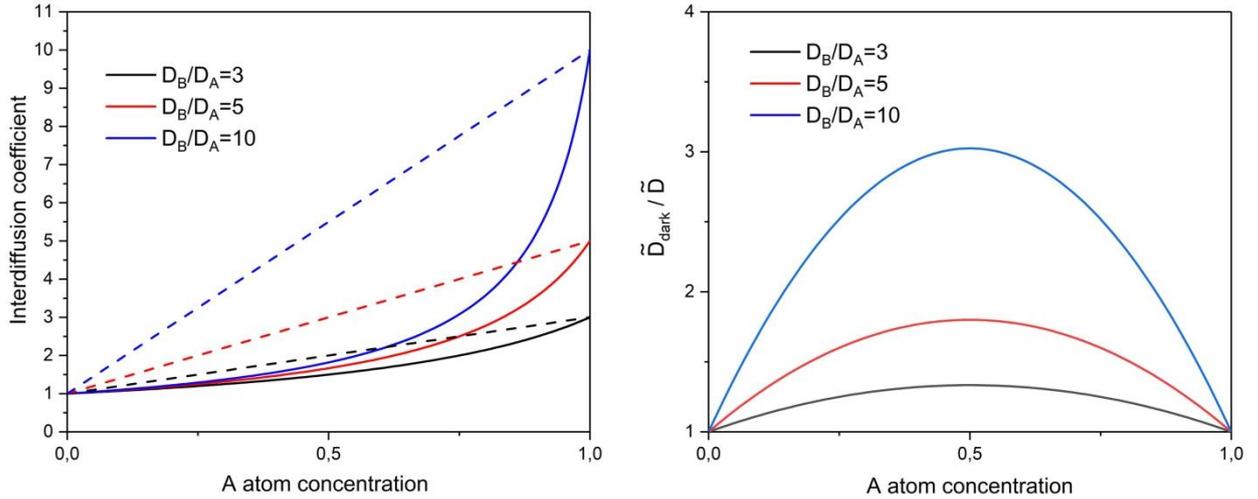

**Fig.1** The composition dependencies of the interdiffusion coefficients at various ratios of the selfdiffusion coefficients. Eq. (22) (solid line) and the Darken equation (dash line).

The dependences show that the larger the ratio of self-diffusion coefficients, the greater the difference $\widetilde{D}$ and $\widetilde{D}_{Dark}$.

## 3. Analysis of interdiffusion in multicomponent alloys.

The above solution approach can be generalized to the many component system ($\alpha = 1, \ldots \ldots, m$). We apply linearization in the same way as it is done for a binary system:

$$c_\alpha(x + \xi, t), \quad c(x + \xi, t); \quad c_\alpha(x + \xi, t) = c_\alpha^0(x, t) + c_\alpha^1(\xi, t), \quad 0 \leq \xi \leq l$$

$$c_v(x + \xi, t) = c^0(x, t) + c^1(\xi, t); \qquad (23)$$

$$\frac{c_\alpha^1}{c_\alpha^0} \ll 1, \quad \frac{c^1}{c^0} \ll 1.$$

Then the linearized equations for the alloy components have the form:

$$\frac{\partial c_\alpha}{\partial \tau} - Z_\alpha c^0 \frac{\partial^2 c_\alpha}{\partial \xi^2} = -Z_\alpha c_\alpha^0 \frac{\partial^2 c_v}{\partial \xi^2}, \qquad (24)$$

$$Z_\alpha = \gamma a^2 \Gamma_\alpha,$$

$$D_\alpha(t) \equiv Z_\alpha c^0(x, t).$$

And for the concentration of vacancies we get:

$$\frac{\partial c_v}{\partial \tau} - D_V(t) \frac{\partial^2 c_v}{\partial \xi^2} = -\sum_\alpha D_\alpha(t) \frac{\partial^2 c^\alpha}{\partial \xi^2}, \qquad (25)$$

where $D_V(t) = \sum_\alpha Z_\alpha c_\alpha^0(x, t)$

After transformations similar to those used earlier when considering a binary system, for slowly varying coefficients of the series of type Eq. (13), which determine the concentration of vacancies, we have the equation:

$$\bar{\varphi}_n(t) = \bar{v}_{12}(t) + \sum_\alpha \frac{D_\alpha(t)}{D_V(t)} \bar{\varphi}_n^\alpha(t) \qquad (26)$$

In this equation, as before, we restrict ourselves only to terms of the first order of the series in the vacancy concentration.

Then Eq. (26) should be substituted into the equations for the fluxes of components, and after transformations similar to Eq. (18), we get:

$$j_\alpha = -\frac{D_\alpha}{\sum_\alpha D_\alpha c_\alpha^0} \sum_\beta D_\beta \left( c_\beta^0 \frac{\partial c_\alpha}{\partial \xi} - c_\alpha^0 \frac{\partial c_\beta}{\partial \xi} \right) + Z_\alpha c_\alpha^0 \frac{c_1^e - c_2^e}{l}, \qquad (27)$$

where $\beta = 1, \ldots, m$ as well as $\alpha$. The condition $\alpha \neq \beta$ is not necessary since with the equality of these indexes the corresponding term in Eq. (27) is reset to zero.

In the case of a three-component system for fluxes, we obtain

$$j_1 = -\frac{D_1 D_2}{D_1 c_1^0 + D_2 c_2^0 + D_3 c_3^0} \left( c_2^0 \frac{\partial c_1}{\partial \xi} - c_1^0 \frac{\partial c_2}{\partial \xi} \right) - \frac{D_1 D_3}{D_1 c_1^0 + D_2 c_2^0 + D_3 c_3^0} \cdot \left( c_3^0 \frac{\partial c_1}{\partial \xi} - c_1^0 \frac{\partial c_3}{\partial \xi} \right) +$$

$$+ Z_1 c_1^0 \frac{c_1^e - c_2^e}{l}. \qquad (28)$$

Or in a accepted view:

$$j_1 = -\frac{c_2^0 D_1 D_2 + c_3^0 D_1 D_3}{D_1 c_1^0 + D_2 c_2^0 + D_3 c_3^0} \frac{\partial c_1}{\partial \xi} + \frac{c_1^0 D_1 D_2}{D_1 c_1^0 + D_2 c_2^0 + D_3 c_3^0} \frac{\partial c_2}{\partial \xi} + \frac{c_1^0 D_1 D_3}{D_1 c_1^0 + D_2 c_2^0 + D_3 c_3^0} \frac{\partial c_3}{\partial \xi} +$$

$$+ Z_1 c_1^0 \frac{c_1^e - c_2^e}{l}. \qquad (29)$$

The equation for the fluxes of components 2 and 3 can be obtained by replacing the indexes. If we carry out transformations taking into account conditions Eq. (20), then we can write for alloys closed to ideal:

$$j_1 = -\frac{c_2 D_1 D_2 + c_3 D_1 D_3}{D_1 c_1 + D_2 c_2 + D_3 c_3} \frac{\partial c_1}{\partial x} + \frac{c_1 D_1 D_2}{D_1 c_1 + D_2 c_2 + D_3 c_3} \frac{\partial c_2}{\partial x} + \frac{c_1 D_1 D_3}{D_1 c_1 + D_2 c_2 + D_3 c_3} \frac{\partial c_3}{\partial x}. \qquad (30)$$

Usually the calculations of the matrix of diffusion coefficients in ternary systems are based on the concentration profiles obtained in the experiment [3,6]. Form of the used equations is similar to Eq. (30)).

Analytical solution of diffusion equations obtained after substitution the flux (Eq. (29) or Eq. (30)) in the continuity equation, apparently is not possible. However in such diffusion equations there are no rapidly relaxing terms, since their contribution to the solutions for the component concentrations has a higher order in the concentration of vacancies. Taking into account the conditions give a possibility to use numerical methods for solving mentioned equations. Should be reminded that under numerical solution of the diffusion equation system (similar to Eqs. (8), (9)) we have the problem which related to the difference in time steps by several orders in equation for vacancy concentration and equations for component concentrations, because $D_V \gg D_\alpha$.

In the case of a four-component system for fluxes we obtain:

$$j_1 = -\frac{c_2 D_1 D_2 + c_3 D_1 D_3 + c_4 D_1 D_4}{D_1 c_1 + D_2 c_2 + D_3 c_3 + D_4 c_4} \frac{\partial c_1}{\partial x} + \frac{c_1 D_1 D_2}{D_1 c_1 + D_2 c_2 + D_3 c_3 + D_4 c_4} \frac{\partial c_2}{\partial x} +$$

$$+ \frac{c_1 D_1 D_3}{D_1 c_1 + D_2 c_2 + D_3 c_3 + D_4 c_4} \frac{\partial c_3}{\partial x} + \frac{c_1 D_1 D_4}{D_1 c_1 + D_2 c_2 + D_3 c_3 + D_4 c_4} \frac{\partial c_4}{\partial x}. \qquad (31)$$

Similarly, it is easy to write equations for fluxes in alloys with many components.

## 4. Illustrative calculations and discussion of the results.

Figure 2 show the dependences of the elements of the matrix of diffusion coefficients (see Eq. (30)) on the concentrations at the ratios of the coefficients of selfdiffusion for the example selected by the following: $D_1 = 7.$, $D_2 = 3.$, $D_3 = 1$. Moreover, to preserve symmetry in the equations for the mentioned elements in this figure, the dependences are presented on three concentrations and the condition $c_1 + c_2 + c_3 \approx 1$ is not used.

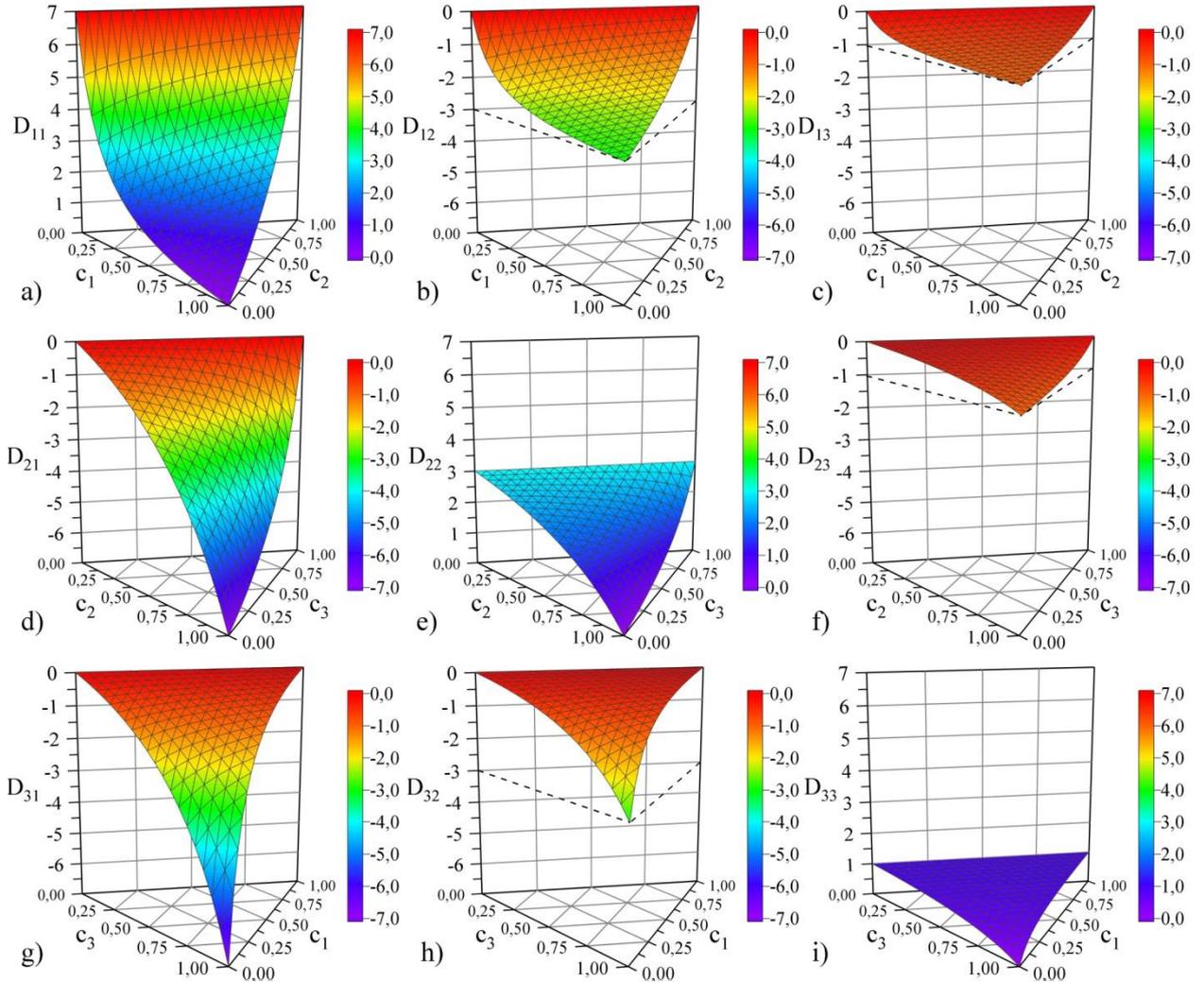

**Fig. 2** Dependence of the elements of the matrix of diffusion coefficients from concentrations

The data in Fig. 2 show that in contrast to Darken theory the elements of the matrix of coefficients of interdiffusion nonlinearly depend on concentrations. The values of these coefficients calculated for concentrations $c_1 = 1/3$, $c_2 = 1/3$, $c_3 = 1/3$ (Table 1).

**Table 1**

| Selfdiffusion coefficients | Interdiffusion coefficients $D_{ij}$ | | |
|---|---|---|---|
| $D_1=7$, $D_2=3$, $D_3=1$ | 2.545 | -1.909 | -0.636 |
|  | -1.909 | 2.182 | -0.273 |
|  | -0.636 | -0.273 | 0.909 |
| $D_1=1$, $D_2=3$, $D_3=7$ | 0.909 | -0.273 | -0.636 |
|  | -0.273 | 2.182 | -1.909 |
|  | -0.636 | -1.909 | 2.545 |
| $D_1=10$, $D_2=5$, $D3=1$ | 3.75 | -3.125 | -0.625 |
|  | -3.125 | 3.4375 | -0.3125 |
|  | -0.625 | -0.3125 | 0.9375 |

As can be seen from these data, the diagonal elements of matrix are significantly smaller than the values of the self-diffusion coefficients and less than diagonal coefficients of Darken theory.

The following are the results of numerical solutions of a system of diffusion equations for a three-component alloy.

## 5. Numerical solutions of the system of diffusion equations.

After substituting the Eqs. (30) for the fluxes into the continuity equations, we obtain a system of three nonlinear diffusion equations. This system is solved for the same values of the selfdiffusion coefficients as above. The calculation results are given on Fig.3.

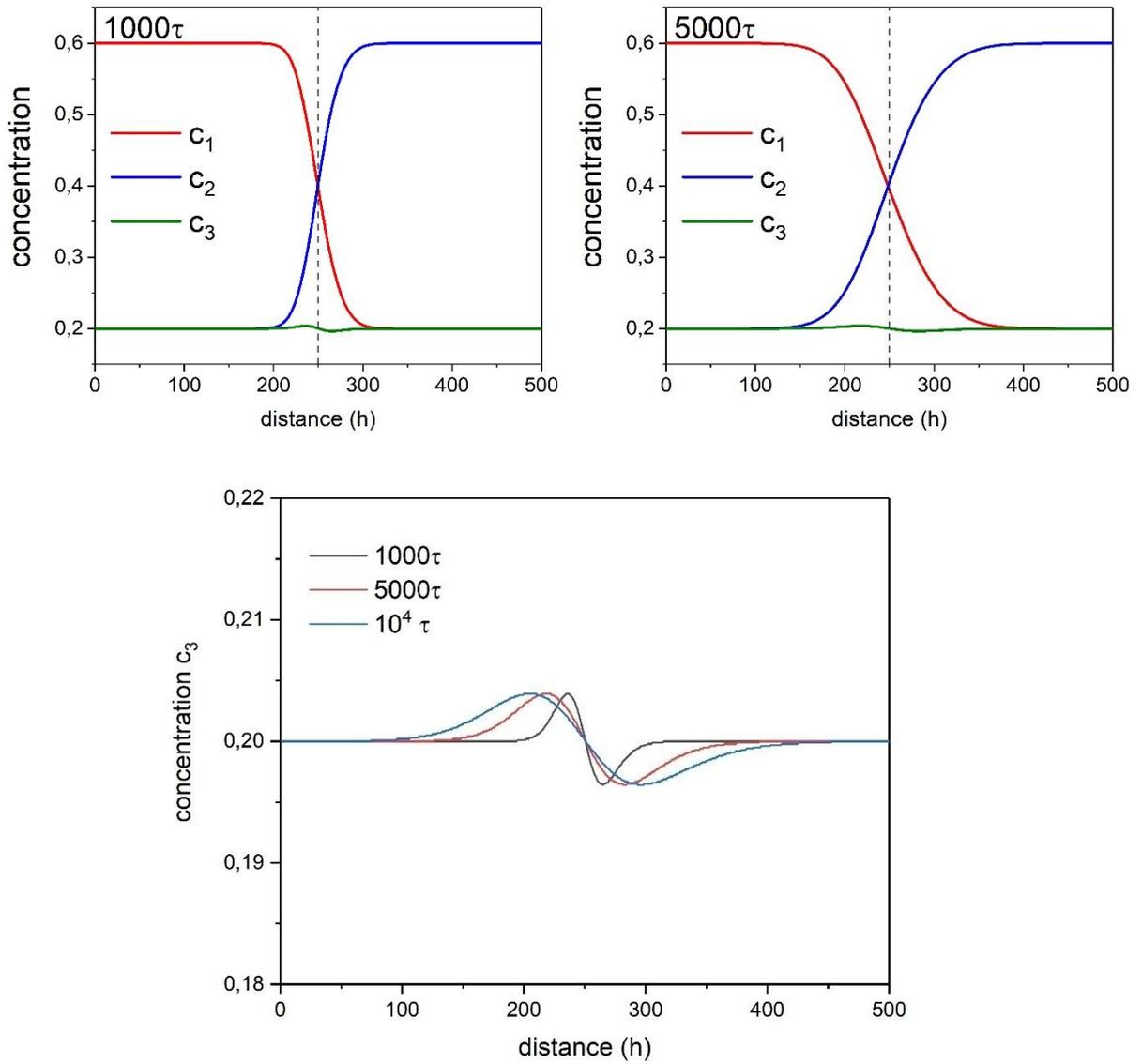

**Fig. 3** Concentration profiles in a three-component system for a different number of time steps ($D_1 = 7$, $D_2 = 3$, $D_3 = 1$)

The following figure shows the calculation results for a different ratio on the coefficients of selfdiffusion.

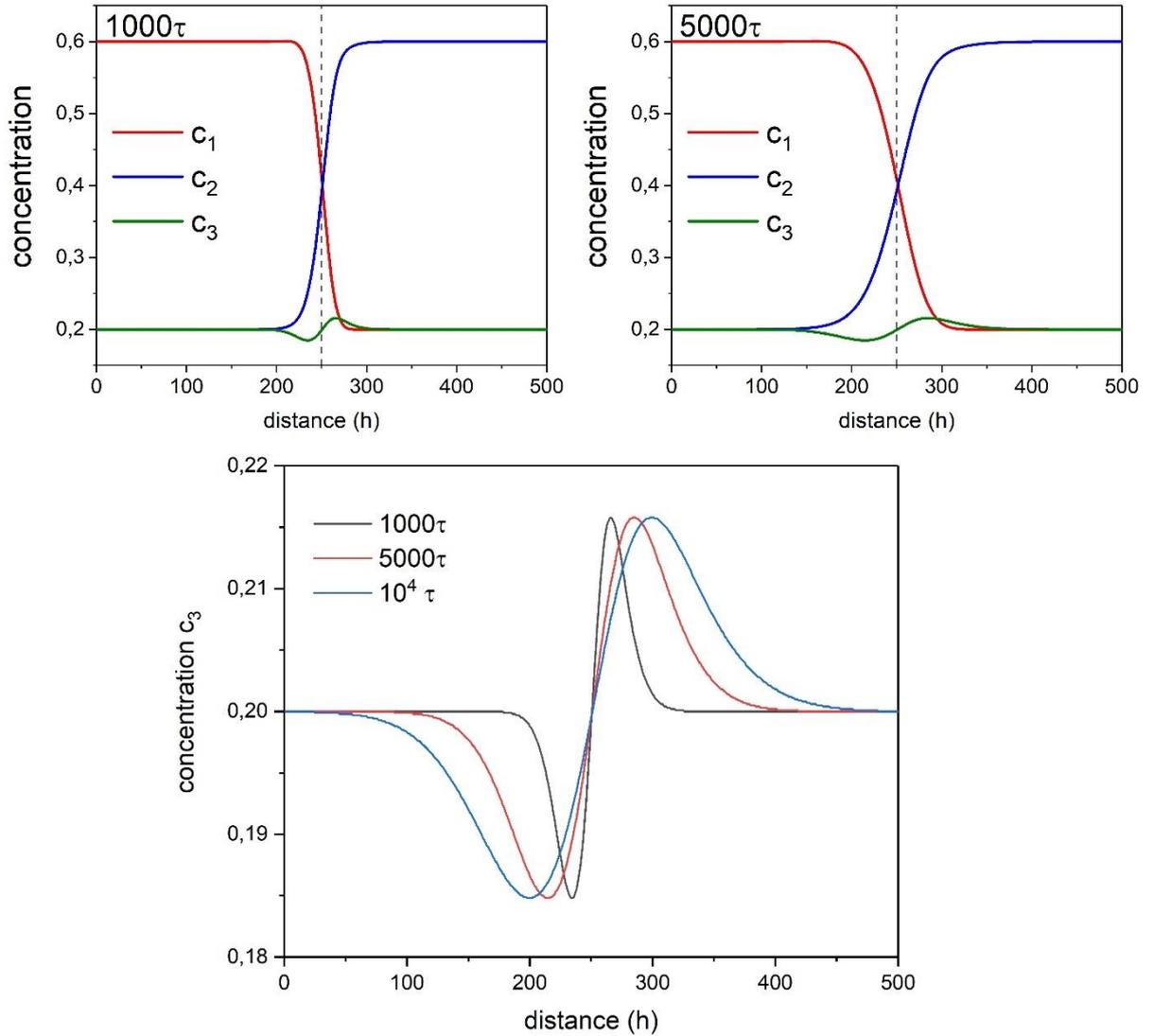

**Fig. 4** Concentration profiles in a three-component system for a different number of time steps ($D_1 = 1$, $D_2 = 3$, $D_3 = 7$)

The data presented in the figures show that if the third component is the least mobile of the components, then its perturbation of the initially constant concentration is very small, and the change in the concentration of the second component would be slower, while in these calculations selfdiffusion coefficient of the second component is the same.

We also note that the calculations performed show the same maximum values of the deviation of the concentration of the third component from the initial one for different times ("undamped" wave process.) In other words, the concentration distribution of the third component has the form of a solitary wave similar to solitons, but this wave not only shifts, but also expands with time.

Fig. 5 and Fig. 6 show the results of similar calculations in the three-component system for equiatomic composition.

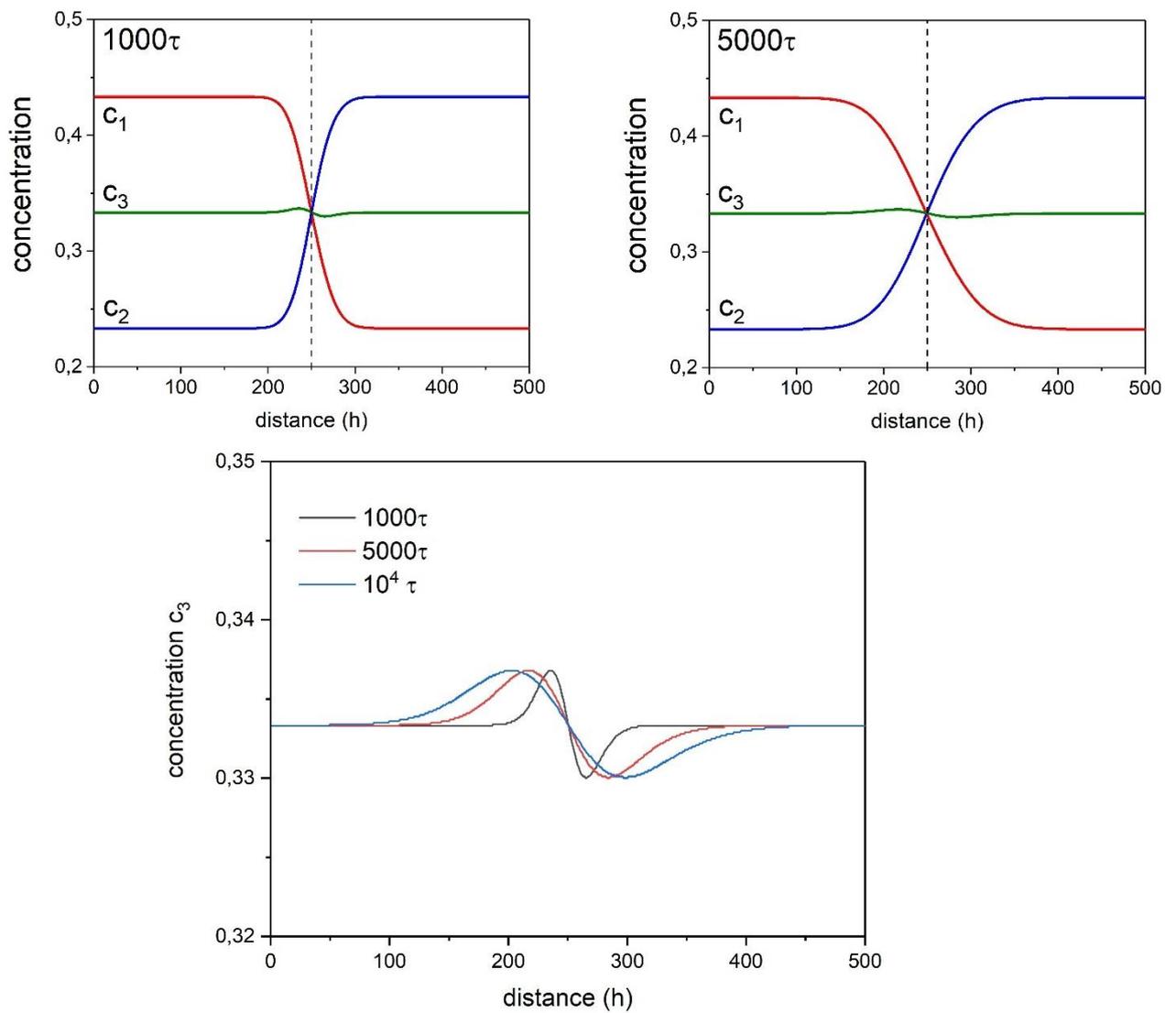

**Fig. 5** Concentration profiles in a three-component system for a different number of time steps for another initial concentrations ($D_1 = 7$, $D_2 = 3$, $D_3 = 1$)

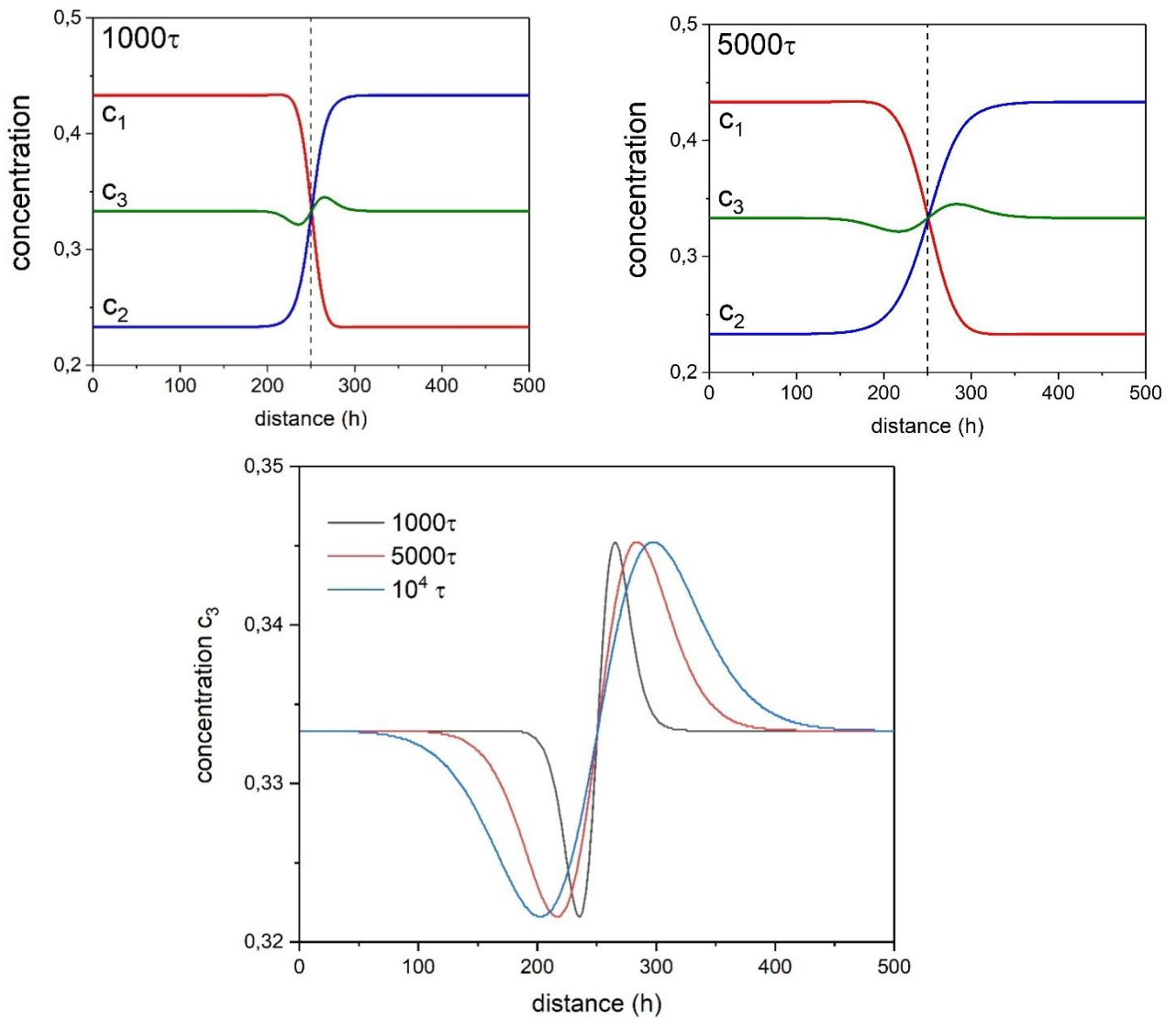

**Fig. 6** Concentration profiles in a three-component system for a different number of time steps for another initial concentrations ($D_1 = 1$, $D_2 = 3$, $D_3 = 7$)

Now we give the values of the elements of the matrix of interdiffusion coefficients in a four-component system for different coefficients of selfdiffusion (Table 2) and equal concentrations (0.25).

Table 2

| Selfdiffusion coefficients | Interdiffusion coefficients $D_{ij}$ | | | |
|---|---|---|---|---|
| $D_1=7, D_2=5, D_3=3, D_4=1$ | 3.9375 | -2.1875 | -1.3125 | -0.4375 |
| | -2.1875 | 3.4375 | -0.9375 | -0.3125 |
| | -1.3125 | -0.9375 | 2.4375 | -0.1875 |
| | -0.4375 | -0.3125 | -0.1875 | 0.9375 |
| $D_1=1, D_2=5, D_3=3, D_4=7$ | 0.9375 | -0.3125 | -0.1875 | -0.4375 |
| | -0.3125 | 3.4375 | -0.9375 | -2.1875 |
| | -0.1875 | -0.9375 | 2.4375 | -1.3125 |
| | -0.4375 | -2.1875 | -1.3125 | 3.9375 |
| $D_1=10, D_2=5, D_3=3, D_4=1$ | 4.7368 | -2.6316 | -1.5789 | -0.5263 |
| | -2.6316 | 3.6842 | -0.7895 | -0.2632 |
| | -1.5789 | -0.7895 | 2.5263 | -0.1579 |
| | -0.5263 | -0.2632 | -0.1579 | 0.9474 |

In conclusion, we present the values of the elements of the matrix of interdiffusion coefficients in a five-component system with different selfdiffusion coefficients (Table 3) and equal concentrations (0.20).

Table 3

| Selfdiffusion coefficients | Interdiffusion coefficients $D_{ij}$ | | | | |
|---|---|---|---|---|---|
| $D_1=7, D_2=6, D_3=5, D_4=3, D_5=1$ | 4.7727 | -1.9091 | -1.5909 | -0.9545 | -0.3182 |
| | -1.9091 | 4.3636 | -1.3636 | -0.8182 | -0.2727 |
| | -1.5909 | -1.3636 | 3.8636 | -0.6818 | -0.2273 |
| | -0.9545 | -0.8182 | -0.6818 | 2.5909 | -0.1364 |
| | -0.3182 | -0.2727 | -0.2273 | -0.1364 | 0.9545 |
| $D_1=1, D_2=6, D_3=5, D_4=3, D_5=7$ | 0.9545 | -0.2727 | -0.2273 | -0.1364 | -0.3182 |
| | -0.2727 | 4.3636 | -1.3636 | -0.8182 | -1.9091 |
| | -0.2273 | -1.3636 | 3.8636 | -0.6818 | -1.5909 |
| | -0.1364 | -0.8182 | -0.6818 | 2.5909 | -0.9545 |
| | -0.3182 | -1.9091 | -1.5909 | -0.9545 | 4.7727 |
| $D_1=10, D_2=6, D_3=5, D_4=3, D_5=1$ | 6.0 | -2.4 | -2.0 | -1.2 | -0.4 |
| | -2.4 | 4.56 | -1.2 | -0.72 | -0.24 |
| | -2.0 | -1.2 | 4.0 | -0.6 | -0.2 |
| | -1.2 | -0.72 | -0.6 | 2.64 | -0.12 |
| | -0.4 | -0.24 | -0.2 | -0.12 | 0.96 |

*Some thoughts about "sluggish" diffusion in systems that have chemical composition inhomogeneity.*

From the results (Fig. 1.2.3). and the ATID equations it follows that, just as with interdiffusion in binary systems, manycomponent diagonal coefficients are less than selfdiffusion coefficients and less than diagonal coefficients of Darken theory. This is one of the reasons for slowing diffusion, which follows from the analysis within ATID.

The second significant factor, in contrast to the binary system, is that in manycomponent alloys the slowing down can be caused not only by smaller diffusion coefficients, but also, as follows from the equations for the fluxes, by the component concentration gradient directionality. Consider, for example, an equation for the flux $j_1$ in a four-component system of equiatomic composition. Suppose for simplifying that the concentration gradients for all components are equal in magnitude. It is easy to see, taking into account the values of the coefficients of the interdiffusion matrix (Table 2), that if the gradient of the second component is parallel to the gradient of the first, then we have a slowing down of the flux, and if the gradient of the third component is parallel to the gradient of the first, then increasing of the flux (while the gradients of the other two components are antiparallel to the gradient of the first component). It is obvious that for some gradients of the second component and the corresponding gradients of the third and fourth, the equality of flux to zero is possible. It can be concluded that, in contrast to the binary system, in multicomponent alloys additional degrees of freedom appear and the slowing down or increasing of diffusion processes will depend not only on the coefficients of the matrix, but also on the geometry of the initial concentration distribution (from the initial and boundary conditions for the system of diffusion equations).

We believe that the factors mentioned above, acting together or separately, can slow down interdiffusion in highly entropy alloys, but the rate of slowing down can apparently be quite different.

**Conclusions**

• The mathematical apparatus of the alternative theory of interdiffusion for multicomponent systems close to ideal solid solutions has been developed.

• Equations for the fluxes of components for multicomponent systems and corresponding matrices of coefficients of interdiffusion are obtained.

• It is shown that consistently taking into account the deviation of the concentration of vacancies from the equilibrium one, which underlies the developed approach, makes it possible to explain the possibility of sluggish diffusion in multicomponent alloys under certain conditions and to identify factors contributing to the slowdown.

**Acknowledgements**

Authors would like to acknowledge the financial support of the National Research Nuclear University MEPhI Academic Excellence Project (Contract No. 02.a03.21.0005).

**Appendix I**

***Solution of the diffusion equation with a time-dependent diffusion coefficient***

$$\frac{\partial c}{\partial t} - D(t)\frac{\partial^2 c}{\partial \xi^2} = f(x,t), \tag{I.1}$$

where $f(x,t)$ is abbreviated notation of the right-hand side of any parabolic equation.

Make a change of variables (this method is described for example in [19]).

$$\tau = \int_0^t \frac{D(t')dt'}{\mathcal{D}}, \quad \text{were } \mathcal{D} = const, \quad t = \Psi(\tau), \quad \mathcal{D}dt = D(t)dt, \tag{I.2}$$

$$\Psi\left(\int_0^t \frac{D(t')dt'}{\mathcal{D}}\right) = t, \quad f(x,t) = f(x,\Psi(\tau)) = \tilde{f}(x,\tau).$$

Then, after the transformations

$$\frac{\partial c}{\partial t} = \frac{\partial \tilde{c}}{\partial \tau}\frac{\partial \tau}{\partial t} = \frac{\partial \tilde{c}}{\partial \tau}\frac{\widetilde{D}(\tau)}{\mathcal{D}} \tag{I.3}$$

$$\frac{\partial \tilde{c}}{\partial \tau} - \mathcal{D}\frac{\partial^2 \tilde{c}}{\partial x^2} = \frac{\tilde{f}(x,\tau)}{\widetilde{D}(\tau)}\mathcal{D}.$$

As a result, we obtain an inhomogeneous equation with a constant diffusion coefficient. The solution can be written using the Green's functions:

$$\tilde{c}(x,\tau) = \sum_{n=1}^{\infty} \varphi_n(0) exp(-\lambda_n^2 \mathcal{D}\tau) sin(\lambda_n x) + \qquad (I.4)$$

$$+ \frac{2\pi \mathcal{D}}{l^2} \sum_{n=1}^{\infty} n \cdot sin(\lambda_n x) \int_0^{\tau} exp\left(-\lambda_n^2 \mathcal{D}(\tau - \tau')\right) \tilde{u}_{12}(\tau') d\tau' +$$

$$+ \sum_{n=1}^{\infty} sin(\lambda_n x) \int_0^{\tau} d\tau' \tilde{f}_n(\tau') \frac{\mathcal{D}}{\widetilde{D}(\tau')} exp(-\lambda_n^2 \mathcal{D}(\tau - \tau')),$$

where $\lambda_n = \frac{\pi n}{l}$.

After that we will move to the initial variable $t$:

$$c(x,t) = \sum_{n=1}^{\infty} \varphi_n(0) exp\left(-\lambda_n^2 \int_0^t D(t') dt'\right) sin(\lambda_n x) + \qquad (I.5)$$

$$+ \frac{2\pi}{l^2} \sum_{n=1}^{\infty} n sin(\lambda_n x) exp\left(-\lambda_n^2 \int_0^t D(t') dt'\right) \int_0^t exp\left(\lambda_n^2 \int_0^{t'} D(t'') dt''\right) D(t') u_{12}(t') dt' +$$

$$+ \sum_{n=1}^{\infty} sin(\lambda_n x) \int_0^t dt' f_n(t') exp\left(-\lambda_n^2 \int_{t'}^t D(t'') dt''\right).$$

***Transformation\****

Substituting $\varphi_n^A$ and $\varphi_n^B$ in (16) makes it necessary to estimate the integrals:

$$Int = \int_0^t \lambda_n^2 D_A(t') \varphi_n^A(t') exp\left(\lambda_n^2 \int_0^{t'} D_V(t'') dt''\right) dt', \qquad (I.6)$$

that we first slightly transform

$$Int = \int_0^t \frac{D_A(t') \varphi_n^A(t')}{D_V(t')} exp\left(\lambda_n^2 \int_0^{t'} D_V(t'') dt''\right) \lambda_n^2 D_V(t') dt', \qquad (I.7)$$

and then we use the method of integration by *parts*:

$$dV = exp\left(\lambda_n^2 \int_0^{t'} D_V(t'') dt''\right) \lambda_n^2 D_V(t') dt', \qquad V = exp\left(\lambda_n^2 \int_0^{t'} D_V(t'') dt''\right), \qquad (I.8)$$

$$dU = \frac{\partial}{\partial t'}\left(\frac{D_A(t')}{D_V(t')} \varphi_n^A(t')\right) dt' = \left(\frac{D_A(t')}{D_V(t')}\right)' \varphi_n^A(t') dt' + \frac{D_A(t')}{D_V(t')} \frac{\partial \varphi_n^A(t')}{\partial t'} dt'$$

$$Int = \left(\left(\frac{D_A(t')}{D_V(t')}\right) \varphi_n^A(t') exp\left(\lambda_n^2 \int_0^{t'} D_V(t'') dt''\right)\right)\Bigg|_0^t - \qquad (I.9)$$

$$- \int_0^t \left(\frac{D_A(t')}{D_V(t')} \varphi_n^A(t')\right)' exp\left(\lambda_n^2 \int_0^{t'} D_V(t'') dt''\right) dt'$$

$$\text{Int} = \left(\left(\frac{D_A(t')}{D_V(t')}\right)\varphi_n^A(t')\exp\left(\lambda_n^2\int_0^{t'}D_V(t'')dt''\right)\right)\Big|_0^t - \qquad (I.10)$$

$$-\left(\frac{1}{\lambda_n^2 D(t')}\left(\frac{D_A(t')}{D_V(t')}\varphi_n^A(t')\right)'\exp\left(\lambda_n^2\int_0^{t'}D_V(t'')dt''\right)\right)\Big|_0^t +$$

$$+\int_0^t\left(\frac{1}{\lambda_n^2 D_V(t')}\left(\frac{D_A(t')}{D_V(t')}\varphi_n^A(t')\right)'\right)'\exp\left(\lambda_n^2\int_0^{t'}D_V(t'')dt''\right)dt'.$$

Next, we take into account that

$$\frac{D_A}{D_V} \sim c^0 \ll 1.$$

$\varphi_n, \varphi_n^A, \varphi_n^B$ can be represented as a sum of slowly and rapidly varying terms:

$$\varphi_n = \bar{\varphi}_n + \hat{\varphi}_n, \qquad \varphi_n^A = \bar{\varphi}_n^A + \hat{\varphi}_n^A, \qquad \varphi_n^B = \bar{\varphi}_n^B + \hat{\varphi}_n^B \qquad (I.11)$$

and if we are only interested in slowly varying terms, it is easy to see that in the integral equation for $\bar{\varphi}_n$ the integral term is transformed using (*) into a series in powers of $c^0$.

If taking into the account only first nonvanishing term of the series, we obtain:

$$\bar{\varphi}_n(t) = \bar{V}_{12}(t) + \frac{D_A(t)}{D(t)}\bar{\varphi}_n^A(t) + \frac{D_B(t)}{D(t)}\bar{\varphi}_n^B(t) \qquad (I.12)$$